\documentclass[
onecolumn,
aps,
prd,
amsmath,
amssymb,
nofootinbib
]{revtex4-2}

\usepackage{graphicx}
\usepackage{dcolumn}
\usepackage{bm}
\usepackage{slashed}
\usepackage{hyperref}
\usepackage{xcolor}
\usepackage{subcaption} 
\usepackage[normalem]{ulem}

\begin{document}


\title{\textbf{Negative diffusion in the Functional Renormalization Group flow for the Quark-Diquark Model} 
}%

\author{Johannes Poeplau}
\email{poeplau@itp.uni-frankfurt.de}
\affiliation{Institut f\"ur Theoretische Physik, Goethe University,
Max-von-Laue-Straße 1, D-60438 Frankfurt am Main, Germany}
\author{Ashutosh Dash}%
 \email{dash@itp.uni-frankfurt.de}
\affiliation{Institut f\"ur Theoretische Physik, Goethe University,
Max-von-Laue-Straße 1, D-60438 Frankfurt am Main, Germany}

\author{Dirk H.~Rischke}
\email{drischke@itp.uni-frankfurt.de}
\affiliation{Institut f\"ur Theoretische Physik, Goethe University,
Max-von-Laue-Straße 1, D-60438 Frankfurt am Main, Germany}
\affiliation{
	Helmholtz Research Academy Hesse for FAIR, Campus Riedberg, Max-von-Laue-Straße 12, D-60438 Frankfurt am Main, Germany
}

\date{\today}

\begin{abstract} 
We investigate the Quark-Diquark Model (QDM) with the Functional Renormalization Group (FRG)  in the Local Potential Approximation. 
In a recent work [arXiv:2510.01066 [hep-ph]], problems were reported in applying this method at low temperatures and large quark chemical potentials, which were attributed to numerical artifacts.
In this work, we trace the origin of these problems to the occurrence of a negative diffusion coefficient during the FRG flow of the derivative of the effective potential, leading to strong oscillations of the latter quantity. 
We show that negative diffusion is a model feature and not a numerical artifact.
We propose a regularization scheme which introduces a hyperdiffusion term to remove these oscillations and demonstrate its effectiveness for studies of the phase diagram of the QDM. 
Regularizing the negative diffusion in the FRG flow is particularly important to reliably study phenomena such as color superconductivity and inhomogeneous phases, which might emerge in the high-density, low-temperature region of the QCD phase diagram.
\end{abstract}

\maketitle


\section{\label{sec:level1}Introduction}

Resolving the phase structure of Quantum Chromodynamics (QCD) at high densities is important to understand the static and dynamical properties of neutron stars and their binary mergers~\cite{PhysRevD.67.074024,Most:2018eaw,baym2018hadrons,Annala_2020}, as well as the dynamics of heavy-ion collisions in the RHIC-BES and FAIR energy range~\cite{Chen:2024aom,Messchendorp:2025men}. 
It is known from perturbative calculations that, at high densities, due to attractive quark-quark interactions, quarks can form Cooper pairs and condense into a BCS-type color-superconducting phase \cite{bailin1979superfluid,Schafer:1999jg,Pisarski:1999tv,rischke2004quark,buballa2005njl,Alford:2007xm}. 
However, even the most extreme densities that occur in Nature, e.g., in the core of neutron stars or in neutron-star mergers, still lie far below the perturbative scale~\cite{gorda2026abovebelowassessingextremeness}.
Furthermore, lattice-QCD calculations  at nonzero net baryon densities are hampered by the fermion sign problem\footnote{Due to this problem, standard Monte Carlo methods, which use importance sampling, fail. Alternative approaches exist~\cite{Philipsen:2021vgp}, but are still restricted to unphysical parameter regions.}.
Therefore, as of yet no first-principle results exist in the range of densities and temperatures relevant for neutron stars or neutron-star mergers.

Dyson-Schwinger equations~\cite{Roberts:1994dr,Roberts:2000aa,Fischer:2018sdj} and the Functional Renormalization Group (FRG)~\cite{wetterich1993exact,Pawlowski:2005xe,Dupuis:2020fhh} are nonperturbative methods which are not impaired by the fermion sign problem and could thus in principle give access to the range of temperatures and densities relevant for neutron stars and their mergers.
In the past, FRG studies of QCD have made remarkable progress in the vacuum as well as for finite temperature and moderate density, where they are in good agreement with results from lattice QCD~\cite{Braun:2017srn,Fu:2019hdw}. Furthermore, advances have been made towards the high-density sector~\cite{Braun:2019aow,pawlowski2025inhomogeneousinstabilitieshighdensityqcd,Sattler:2025hcg,Fu:2024rto}.
However, in order to obtain conclusive results, one would have to account for the various phenomena that may occur in this region, such as the moat regime, inhomogeneous phases, and diquark condensation, which presents various technical problems when one tries to include them in the FRG framework.
Therefore, as a first step, most FRG studies of the high-density region have simplified the problem by studying effective models for the strong interaction instead of QCD itself~\cite{Schaefer_2005,Pawlowski:2014zaa,Tripolt_2014,Ihssen:2023xlp,Tripolt:2021jtp,Otto:2020hoz}.

However, in the high-density, low-temperature region of the phase diagram of strong-interaction matter, also these models exhibit problems.
For instance, the Quark-Meson Model (QMM) was found to have negative entropy density in parts of that region~\cite{Tripolt_2018}.
It was speculated that accounting for diquark condensation could cure this model artifact~\cite{Sedrakian:2017qpg}, which suggests extending the QMM by diquark degrees of freedom.
However, investigations in the framework of the Quark-Diquark Model (QDM) and the Quark-Meson-Diquark Model (QMDM)~\cite{Braun:2018svj,lakaschus2021inhomogeneous,stoll2025nonperturbative,andersen2024colorsuperconductivityspeedsound,Gholami:2026cgd,Mire:2026auc} were not able to provide a clear picture of the phase structure at low temperatures and high densities, at least not beyond the mean-field approximation, because these studies were either conducted at very low or zero chemical potential~\cite{Gholami:2026cgd}, used a low numerical resolution~\cite{lakaschus2021inhomogeneous}, or observed oscillations~\cite{stoll2025nonperturbative} during the flow\footnote{Oscillations can occur when the effective potential is discretized on a grid, as, e.g., in Refs.~\cite{stoll2025nonperturbative,lakaschus2021inhomogeneous}, while they are absent for a Taylor expansion of the effective potential. The reason is that, for a Taylor expansion, the resulting set of equations are ordinary differential equations instead of partial differential equations (PDEs) and our findings only apply to the PDE formulation of the flow equations.}.
Oscillations may also arise in other models using complex scalar fields -- such as those to describe Bose-Einstein condensation or pion condensation -- since the corresponding flow equations share structural similarities with the QDM and QMDM~\cite{Terazaki:2024evv,Svanes_2011}.

In this work, we study the QDM with the FRG in the Local Potential Approximation (LPA). 
The QDM is a simpler version of the QMDM: it only contains quark and diquark degrees of freedom, but no scalar and pseudoscalar mesons.
It is thus not able to describe chiral symmetry breaking in the low-temperature, low-density region of the QCD phase diagram, but it is well equipped to study diquark condensation at low temperatures and moderate to high densities.
Furthermore, studying the QMDM within the FRG framework is computationally highly demanding, because it requires two dimensions in field space -- one for the chiral order parameter and one for the order parameter for diquark condensation.
In contrast, the QDM only requires a single dimension in field space and the corresponding FRG flow equations have a much simpler structure. 

A fairly new development in the application of the FRG is the realization that the corresponding flow equations can be cast into the form of a fluid-dynamical advection-diffusion equation for the derivative of the effective potential~\cite{Grossi:2019urj,Koenigstein:2021syz}.
A recent study of the QDM applied such a fluid-dynamical framework~\cite{stoll2025nonperturbative} and observed oscillations in the derivatives of the effective potential at low temperatures and high chemical potential, which were attributed to numerical artifacts. 
In this paper, we analyze this phenomenon in more detail and show that these oscillations arise from a negative diffusion coefficient during the FRG flow. 

From the theory of PDEs it is well known that negative diffusion causes the diffusion equation to be ill-posed and, in general, not solvable with standard analytical or numerical methods~\cite{Tikhonov1995}.
Nevertheless, a well-known method~\cite{witelski1995shocks,miller2025shock,Maron:2008sb} to circumvent the issue of ill-posedness and render the problem solvable is to add a regularizing higher-derivative term to the flow equations, which converts the flow equation into a well-posed Cahn-Hilliard equation~\cite{CahnHilliard}, thus eliminating the oscillations.
Here, we explicitly demonstrate the viability of this so-called \textit{hyperdiffusion method}~\cite{Maron:2008sb} and present, for the first time, a phase diagram for the QDM which is reliable also in the low-temperature, high-density region.

This paper is organized as follows.
In Sec.~\ref{sec:1} we introduce the QDM and the FRG framework. Afterwards, in Sec.~\ref{sec:2} we  show how to write the FRG flow equation as an advection-diffusion equation. 
Then, in Sec.~\ref{sec:3} we discuss our results regarding negative diffusion in the QDM and how to obtain reliable results with the hyperdiffusion method.
We conclude this paper in Sec.~\ref{sec:conclusions} with a summary of our results and an outlook.
A concise overview of the numerical details of our calculations and model parameters can be found in the Appendix.

\section{The Quark-Diquark Model with the FRG}\label{sec:1}

The QDM is an effective model for strong-interaction matter at high densities, where the degrees of freedom are quarks and diquarks. 
Diquarks are pairs of two quarks that can form around the Fermi surface at high densities, if there is an attractive quark-quark interaction. 
For an overview of quark matter at high densities and color superconductivity, see Refs.~\cite{bailin1979superfluid,rischke2004quark,buballa2005njl,Alford:2007xm}.

We examine the QDM with $N_f = 2$ flavors and $N_c = 3$ colors in 3+1 dimensions, where diquarks are expected to be in the two-flavor color-superconducting (2SC) phase at high densities. 
We only consider pairing in the fully antisymmetric scalar diquark channel, as it is predicted to be the most dominant one~\cite{buballa2005njl}. 
In the 2SC phase, the QDM has the following Euclidean action~\cite{Shu2009wENO}
\begin{eqnarray}
	S &= & \int_0^{1/T}\mathrm{d} \tau\,\int_V \mathrm{d}^3x\,\left[\bar{\psi}\left(\slashed{\partial} - \mu \gamma^0\right)\psi + \frac{h}{2}\left(\Delta^*_a\psi^T\epsilon^a\tau^2C\gamma^5\psi+\Delta_a\bar{ \psi}\epsilon^a\tau^2C\gamma^5\bar{\psi}^T\right)\right.\nonumber\\
&&	\hspace*{2.5cm} +  \left.\left(\partial_\mu-2\mu\,\delta_{0\mu}\right)\Delta_a^*\left(\partial^\mu+2\mu\,\delta^{0\mu}\right)\Delta_a + U(\rho)\vphantom{\frac{h}{2}}\right]\;.\label{eq:Lagrangian}
\end{eqnarray}
Here, the quark field is represented by $\psi$ and its Dirac adjoint by $\bar{\psi}$.
The complex diquark fields  $\Delta_a^*$ and $\Delta_a$ can also be written in terms of real fields 
$\Delta_{a,r}$, $\Delta_{a,i}$, such that $\Delta_a  \equiv \frac{1}{\sqrt{2}}(\Delta_{a,r} + i\Delta_{a,i})$, $\Delta_a^* \equiv \frac{1}{\sqrt{2}}(\Delta_{a,r} - i \Delta_{a,i})$. 
The effective potential of the diquarks $U(\rho)$ depends only on the invariant $\rho \equiv \sum\limits_{a}\Delta^*_a\Delta_a = \sum\limits_{a}\frac{1}{2}(\Delta_{a,r}^2+\Delta_{a,i}^2)$. 
The charge-conjugation matrix in Dirac space is $C = \gamma^2\gamma^0$, the second Pauli matrix $\tau^2$ acts in flavor space, and $\epsilon^a,\, a = \{2,5,7\}$ are the antisymmetric Gell-Mann matrices in color space. 
The chemical potential is denoted by $\mu$, the temperature by $T$, and the spatial volume of the system by $V$, while $h$ is the Yukawa-type quark-diquark coupling constant. 

The FRG implements Wilson's idea of integrating out momentum shells   \cite{PhysRevB.4.3174}. 
For this purpose, a regulator $R_k$, which depends on the renormalization-group (RG) scale $k$, is added to the action in the generating functional. 
The regulator acts as a mass term for momenta smaller than $k$, while leaving the high-momentum part of the theory unaffected. 
In this paper, we use the flat Litim regulator~\cite{Litim:2001up} for spatial momenta.

We employ the FRG flow equation in the form suggested by Wetterich ~\cite{wetterich1993exact}, which describes the flow of the generating functional of one-particle irreducible vertices at the scale $k$, the effective average action $\Gamma_k[\varphi]$, with the RG scale:
\begin{equation}\label{eq:Wetterich}
	\partial_t \Gamma_k[\varphi] = \frac{1}{2}\mathrm{STr}\left[\left(\partial_t R_k\right)\left(\Gamma_k^{(2)} + R_k\right)^{-1}\right]\;,
\end{equation}
where the supertrace STr includes a trace over all field indices, integrals over internal-loop momenta, as well as appropriate signs arising from anticommuting fields.
The RG time is defined as $t\equiv -\log(\frac{k}{\Lambda_{\text{UV}}})$ and $\varphi$ is a superfield that contains all fields of the theory. 
For an overview of the Wetterich equation and common truncations, see Refs.~\cite{Dupuis:2020fhh,morris1994truncations,Pawlowski:2005xe}. 
The integration is performed from the ultraviolet (UV) scale $\Lambda_{\text{UV}}$ to the infrared (IR) scale $\Lambda_{\mathrm{IR}}$.
In principle, the latter scale should be chosen as small as possible, since only for $k \to 0$, the regulator vanishes and the effective average action becomes the full quantum effective action of the theory $\lim\limits_ {k\to0}\Gamma_k[\varphi]\to \Gamma[\varphi]$.
In practice, however, one monitors suitable observables (e.g., the minimum of the effective potential or the curvature mass of the non-Goldstone mode)  during the flow and terminates the integration at a sufficiently small, but finite value of $k=\Lambda_{\mathrm{IR}}$, where these observables no longer change within a prescribed numerical error.

Since the effective action is the Legendre transform of the Schwinger functional $W[J]$, it is by definition convex. 
While the scale-dependent effective average action does not need to be convex during the RG flow, in the IR  convexity is restored~\cite{litim2006convexity}.
This property will be of relevance in Sec.~\ref{sec:2B}, where we use it to constrain the possible shapes that the effective potential could assume during the flow.
 
The LPA is a simple, but effective truncation to solve Eq.~\eqref{eq:Wetterich}, where $\Gamma_k[\varphi]$ is evaluated on a constant background $\bar{ \psi}(x) = \psi(x) = 0,\,\, \rho(x)  \equiv \rho$, and the effective potential $U_k(\rho)$ is taken to be the only part of $\Gamma_k[\varphi]$ that depends on $k$. 
For the sake of convenience, the diquark condensate is rotated in color space, so that only the real part of the $a=2$ component has a non-vanishing expectation value $\Delta_{2,r} \equiv \Delta$, while the expectation values of all other components vanish, $\Delta_{2,i} = \Delta_{5,r} = \Delta_{5,i} = \Delta_{7,r} = \Delta_{7,i} =0$.

 The LPA flow equation for the effective potential of the QDM in $3+1$ dimensions can be written as~\cite{Sattler:2025hcg,stoll2025nonperturbative}\footnote{While the equations derived in Ref.~\cite{Sattler:2025hcg} are for the QMDM, they are written in a way that allows us to straightforwardly identify the flow equations of the QDM.}:
\begin{equation}
\label{eq:flow_U}
	\partial_t U_k = Q + F + S\;.
\end{equation}
The term
\begin{equation}
	Q \equiv -\frac{k^5}{12\pi^2}\left[ \left(1+\frac{8\mu^2}{\chi_D}\right)\frac{\coth\left(\frac{\xi^+}{2T}\right)}{\xi^+}+\left(1-\frac{8\mu^2}{\chi_D}\right)\frac{\coth\left(\frac{\xi^-}{2T}\right)}{\xi^-}\right]\label{eq:Diff}
\end{equation}
results from the $\Delta_2$ diquark loop, with the energies
\begin{equation}
\xi^\pm \equiv \sqrt{k^2+4\mu^2+\frac{1}{2}\left(m_\Delta^2+M_\Delta^2\right) \pm \chi_D}\;,\label{eq:Diff_E_pm}
\end{equation}
where
\begin{align}\label{eq:masses}
m_\Delta^2 & \equiv \frac{\partial_\Delta U_k}{\Delta}\;,\quad M_\Delta^2 \equiv \partial_\Delta^2 U_k\;, \quad
\chi_D  \equiv \sqrt{16\mu^2\left[k^2+\frac{1}{2}\left(m_\Delta^2+M_\Delta^2\right)\right]+\frac{1}{4}\left(m_\Delta^2-M_\Delta^2\right)^2}\;.
\end{align}
The remaining diquarks give the following contribution,
\begin{equation}
	F	\equiv -\frac{(N_c-1)k^5}{12\pi^2}\left[\frac{\coth\left(\frac{\sqrt{k^2+m_\Delta^2}+2\mu}{2T}\right)}{\sqrt{k^2+m_\Delta^2}} +\frac{\coth\left(\frac{\sqrt{k^2+m_\Delta^2}-2\mu}{2T}\right)}{\sqrt{k^2+m_\Delta^2}}\right]\label{eq:Adv_E_m} \; .    
\end{equation}
The two quarks which couple to the condensing diquark field give the contribution 
\begin{equation}
		S \equiv \frac{N_fk^5}{3\pi^2}\left[\frac{k-\mu}{k}\,\,\frac{\tanh\left(E_q^-/2T\right)}{E_q^-}+\frac{k+\mu}{k}\,\,\frac{\tanh\left(E_q^+/2T\right)}{E_q^+}\right]\;,
        \end{equation}
where the fermionic energies are
\begin{equation}
        E_q^\pm = \sqrt{(k\pm\mu)^2 + \frac{(h\Delta)^2}{2}}\;.
\end{equation}
We omitted the contribution of the third quark, which does not participate in diquark condensation, because it is independent of $\Delta$ and thus does not influence the flow of the derivative of the effective potential.

\section{Fluid-dynamical Formulation of the FRG}\label{sec:2}

It has been found~\cite{Grossi:2019urj,Koenigstein:2021syz} that by taking a field derivative of the flow equation~\eqref{eq:flow_U} for $U_k$ and defining $u \equiv \partial_\Delta U_k$, the structure of the resulting equation,
\begin{equation}
\label{eq:KT}
	\partial_t u(\Delta) = \frac{\mathrm{d}}{\mathrm{d}\Delta} F(u,\Delta) + \frac{\mathrm{d}}{\mathrm{d}\Delta} Q(u,\partial_\Delta u,\Delta) + \partial_\Delta S(\Delta)\;,
\end{equation}
is identical to an advection-diffusion equation in conservative form, with $F$ being the advection flux, $Q$ a diffusion-like term, and $S$ a source term. 
This fluid-dynamical analogue provides a more intuitive understanding of the individual terms in Eq.~\eqref{eq:flow_U} and allows the use of well-established algorithms from fluid dynamics, which are optimized to handle non-linear behavior like shock formation possibly occurring during the flow.

A more careful investigation of the diffusion-like term shows that it contains a flux component as well as a diffusion component:
\begin{equation}
\label{eq:decomp_Q}
	\frac{\mathrm{d}}{\mathrm{d}\Delta} Q(u,\partial_\Delta u,\Delta) =  D(u,\partial_\Delta u, \Delta)\partial^2_\Delta u + E(u,\partial_\Delta u, \Delta) \partial_\Delta u+ G(u,\partial_\Delta u, \Delta)\;,
\end{equation}
with the diffusion coefficient
\begin{equation}
	D(u,\partial_\Delta u, \Delta) \equiv \frac{\partial Q(u,\partial_\Delta u,\Delta)}{\partial (\partial_\Delta u)}\;,\label{eq:Diff_fkt}
\end{equation}
a contribution to the propagation speed
\begin{equation}
	E(u,\partial_\Delta u,\Delta) \equiv \frac{\partial Q(u,\partial_\Delta u,\Delta)}{\partial u}\;,
\end{equation}
and another source term
\begin{equation}
	G(u,\partial_\Delta u,\Delta) \equiv \partial_\Delta Q(u,\partial_\Delta u,\Delta)\;,
\end{equation}
which, when expanded, also contains flux-like contributions.
The details of the numerical solution of Eq.~\eqref{eq:KT} are deferred to Appendix \ref{sec:appendix_A}.

\section{Regularizing Negative Diffusion via Hyperdiffusion}\label{sec:3}
In this section, we discuss negative diffusion in the QDM and how to regularize it with the hyperdiffusion method. 
In the first subsection we describe the effects of negative diffusion using the example of the heat equation and how the resulting ill-posedness is regularized by introducing a hyperdiffusion term into the equation. 
In the next subsection, we show both numerically and analytically that negative diffusion can occur during the FRG flow of the derivative of the effective potential of the QDM and can cause oscillations. 
Furthermore, we demonstrate that regularization with the hyperdiffusion method can remove them. 
Finally, we show that negative diffusion can in principle affect a large part of the low-$T$, high-$\mu$ region of the phase diagram of the QDM.
Therefore, employing a regularization such as the hyperdiffusion method is indispensable to obtain reliable results.

\subsection{Negative Diffusion in the Heat Equation}
While uncommon in physics, negative diffusion can appear for instance in biology to model population dynamics, when populations locally concentrate instead of dispersing~\cite{grindrod1988models,grunbaum1994modelling}. 
Negative diffusion is thus a model feature, and not necessarily a numerical artifact.
When the diffusion coefficient turns negative, the initial-value problem becomes ill-posed. 
This means that small perturbations increase exponentially over time and the uniqueness of the solution is not guaranteed \cite{Tikhonov1995}. 
This is especially a problem in numerical simulations, where the solution always contains numerical noise, which, in the case of negative diffusion, can grow quickly and degrade the quality of the solution.

The ill-posedness can be understood by considering the well-known heat equation with a constant diffusion coefficient $D_0$,
\begin{equation}
    \partial_t u = D_0\partial_\Delta^2 u\;.
\end{equation}
We make a plane-wave ansatz, $u(t,\Delta) = \sum_q c_q(t)e^{iq\Delta}$, to solve the equation and look at the evolution of a single mode $c_q(t)$, 
\begin{equation} \label{eq:single_mode}
	\partial_t \, c_q(t) = -D_0 \, q^2\, c_q(t)\;.
\end{equation}
For the standard case of a positive diffusion coefficient, $D_0>0$, all modes are exponentially damped, $c_q(t) \sim \exp(-D_0 q^2 t)$, with a damping rate $D_0 q^2$ which grows quadratically with $q$. 
However, if $D_0$ turns negative, all modes increase exponentially, $c_q(t) \sim \exp(|D_0| q^2 t)$, with high-$q$ modes growing faster than low-$q$ modes. 
Thus, when solving this equation with negative diffusion numerically on a grid with a finite grid spacing, the latter sets the scale for the highest possible $q$, and thus the oscillations will manifest on the scale of the grid spacing.

The problem of ill-posedness can be addressed by adding a regularizing term to the equation and solving it in its weak formulation~\cite{witelski1995shocks,miller2025shock}. 
The regularizing term is chosen such that the solution does not exhibit oscillations caused by negative diffusion. 
In the end, one has to perform the limit of a vanishing regularizing term, in which case the solution of the original equation converges to the correct weak solution.
Of course, numerically, this limit can only be taken up to the point where oscillations spoil the solution, see discussion in Appendix~\ref{sec:appendix_B}.

For our numerical calculations, we regularize the negative diffusion by adding the following hyperdiffusion term~\cite{Maron:2008sb},
\begin{equation}
	R := -C\,\partial_\Delta^4\, u\;,\label{eq:Regularization}
\end{equation}  
with a parameter $C>0$, to the standard diffusion term. 
The single-mode equation~\eqref{eq:single_mode} then becomes
\begin{equation}
	\partial_t c_q(t) = -(D_0 + Cq^2)\,q^2\, c_q(t)\;.
\end{equation}
In case of a negative diffusion coefficient, the regularizing term dampens modes with wave number $q>\sqrt{|D_0|/C}$, thereby eliminating oscillations on a length scale $\Delta \sim 1/q <  \sqrt{C/|D_0|}$.

\subsection{Negative Diffusion in the QDM}\label{sec:2B}

\begin{figure}[htbp!]
  \centering
  \begin{minipage}{0.46\textwidth}
    \centering
    \includegraphics[width=\linewidth]{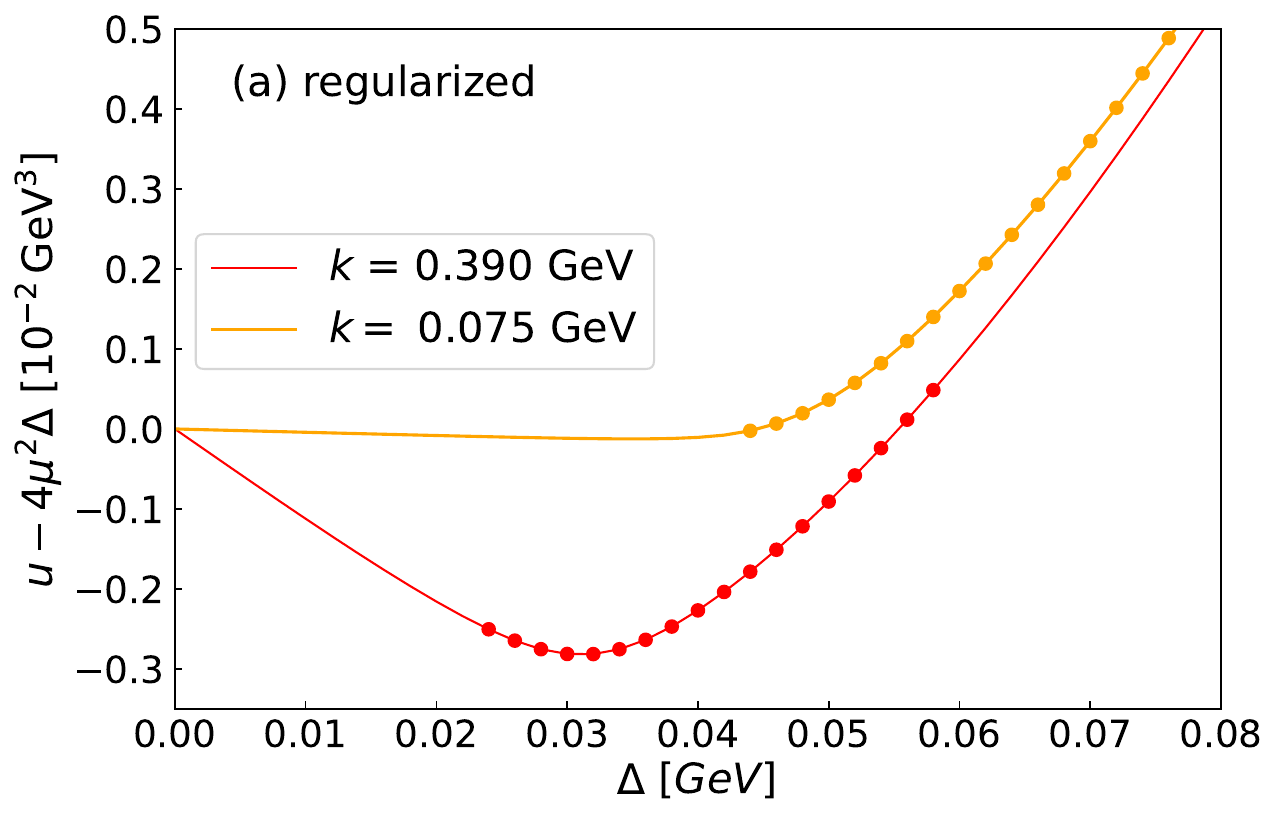}
  \end{minipage}
  \hfill
  \begin{minipage}{0.46\textwidth}
    \centering
    \includegraphics[width=\linewidth]{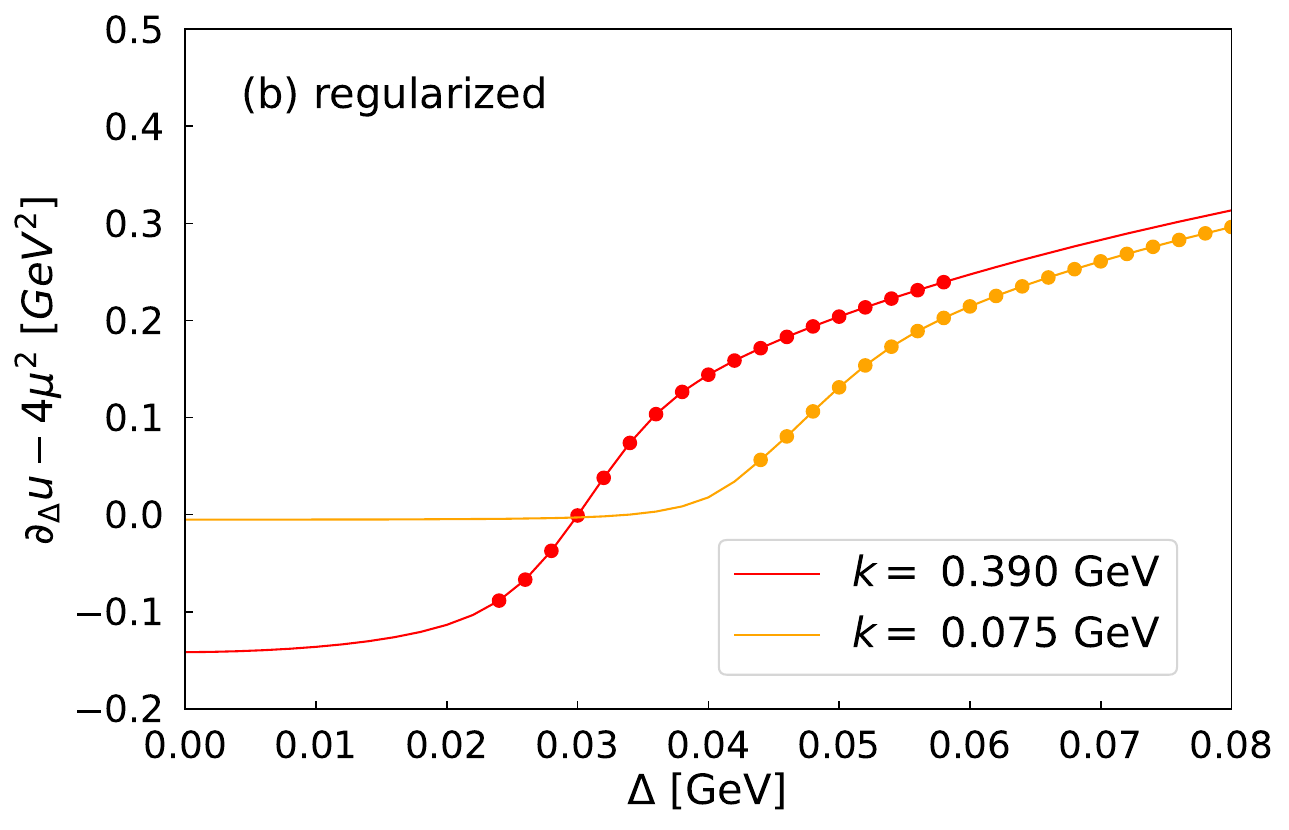}
  \end{minipage}
    \begin{minipage}{0.46\textwidth}
    \centering
    \includegraphics[width=\linewidth]{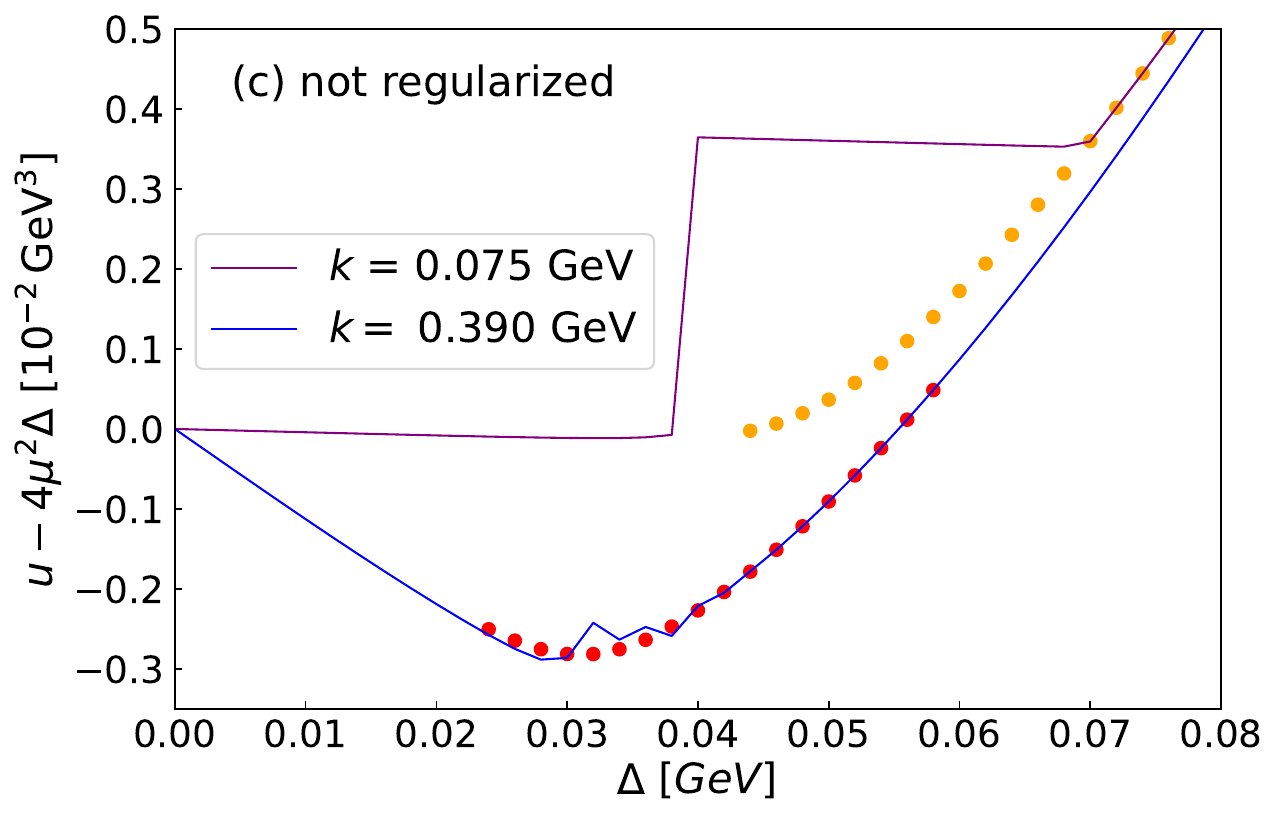}
  \end{minipage}
  \hfill
  \begin{minipage}{0.46\textwidth}
    \centering
    \includegraphics[width=\linewidth]{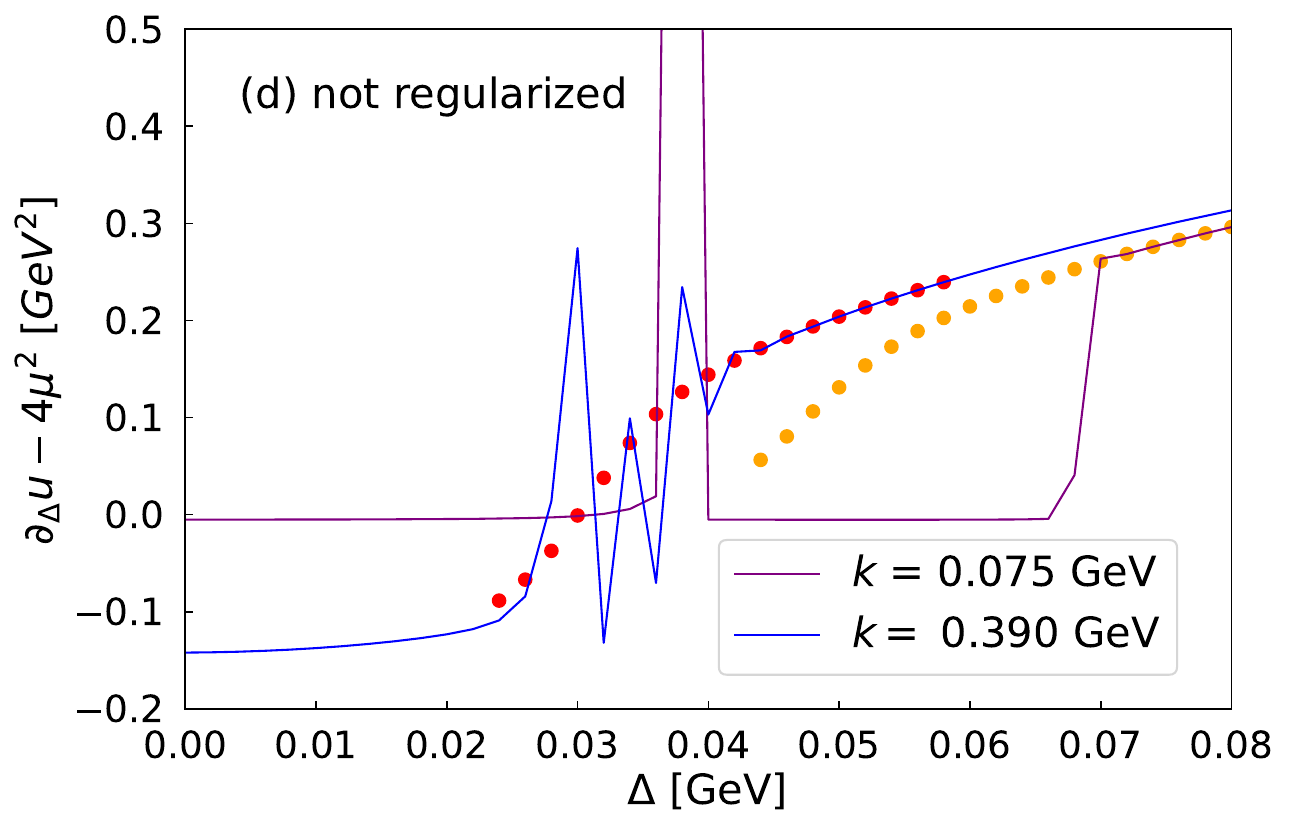}
  \end{minipage}
\caption{\label{fig:comparison} Comparison of (a, c) $u - 4 \mu^2 \Delta$ and (b, d) its first derivative as a function of $\Delta$ for (a, b) the regularized case and (c, d) the not-regularized case, for two exemplary values of $k$ during the FRG flow: the red and blue lines are for $k=0.39\,\text{GeV}$, while the orange and purple lines are for $k=0.075$ GeV. 
We use parameter set 2 in Tab.~\ref{tab:1}, for $T = 0.01\,\text{GeV}$ and $\mu = 0.35\,\text{GeV}$. The colored points in (a, b) indicate where the numerical value of the diffusion coefficient, calculated on the regularized solution, is negative and are overlaid on the not-regularized case (c, d) to guide the eye.
The colored points lie inside the respective negative-diffusion region of Fig.~\ref{fig:regionplot}.}
\end{figure}

\begin{figure}[!htbp]
\hspace*{-1cm}  \begin{minipage}{0.32\textwidth}
    \includegraphics[width=1.2\linewidth]{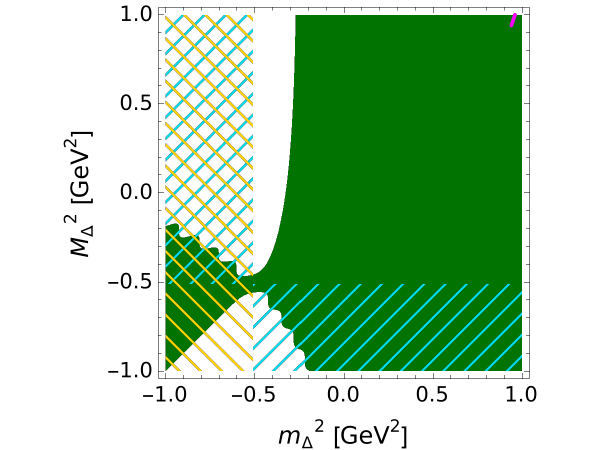}
    \caption*{\hspace{0.34\textwidth}(a)\,\, $k$ = 1.0 GeV}
  \end{minipage}
  \begin{minipage}{0.32\textwidth}
    \includegraphics[width=1.2\linewidth]{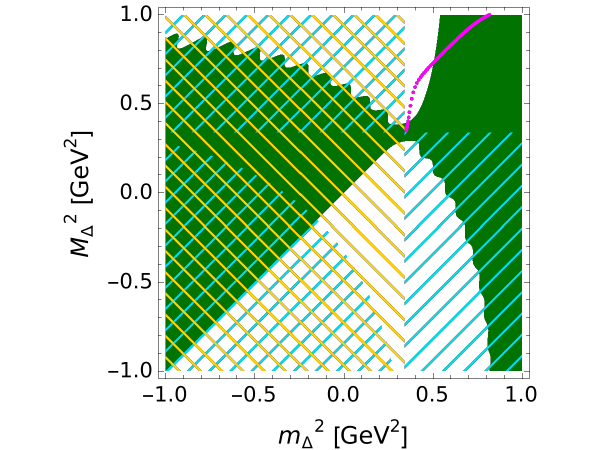}
    \caption*{\hspace{0.37\textwidth}(b)\,\, $k$ = 0.39 GeV}
  \end{minipage}
\begin{minipage}{0.32\textwidth}
    \includegraphics[width=1.2\linewidth]{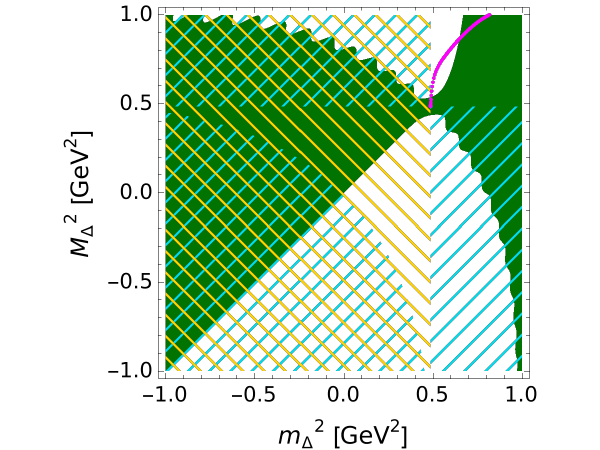}
    \caption*{\hspace{0.4\textwidth}(c)\,\, $k$ = 0.075 GeV}
\end{minipage}
	\caption{\label{fig:regionplot}The $(m_\Delta^2, M_\Delta^2)$ plane for $T=0.01$ GeV and $\mu=0.35$ GeV, at RG scales of (a) $k=1.0$ GeV, (b) $k=0.39$ GeV, and (c) $k=0.075$ GeV. 
     In the green region the diffusion coefficient is positive. The blue-hashed region is excluded by a pole in Eq.~\eqref{eq:Diff} and the orange-hashed region by a pole in Eq.~\eqref{eq:Adv_E_m}. In the white, non-hashed region the diffusion coefficient is negative. The magenta-colored points are the $(m_\Delta^2, M_\Delta^2)$ values calculated from the numerically determined regularized function $u(\Delta)$ shown in Fig.~\ref{fig:comparison} (a)\footnote{The \texttt{Mathematica} notebook used to create these plots can be found in the supplementary material.}.}
\end{figure}

When investigating the QDM, we found that for high chemical potentials and low temperatures, strong oscillations can occur during the flow of $u$, which invalidate the standard numerical method to solve the FRG flow equation~\eqref{eq:KT}.
These oscillations of $u$, which we show in Fig.~(\ref{fig:comparison}) (c) for exemplary values of $k$, become even more pronounced in the derivatives of $u$, as can be seen in Fig.~\ref{fig:comparison} (d).
Although they appear to the left of the physical point (where $u- 4 \mu^2 \Delta =0$), diffusion causes them to propagate into the region around the physical point, making it of paramount importance to understand and eliminate this phenomenon. 

The analogy with the heat equation prompted us to take a closer look at the diffusion coefficient~(\ref{eq:Diff_fkt}), which we determined numerically.
We observed that the oscillations appear at an RG time when the diffusion coefficient becomes negative.
For a better understanding, we computed the diffusion coefficient as a function of $m_\Delta^2$ and $M_\Delta^2$.
This can be done purely analytically, since from Eqs.~\eqref{eq:Diff}, \eqref{eq:Diff_E_pm}, \eqref{eq:masses}, and \eqref{eq:Diff_fkt} one observes that, for given $k$, $T$, and $\mu$, these two parameters capture the complete dependence of the diffusion coefficient on the effective potential and its derivatives, respectively.
Thus, mapping the $(m_\Delta^2, M_\Delta^2)$ plane captures all possible shapes of $u$. 
We choose values of $m_\Delta^2$ and $m_\Delta^2$ in the interval $[-1,1]\,\mathrm{GeV}^2$, as this represents the range of values most relevant for our study.

In the green areas in Figs.~\ref{fig:regionplot} the diffusion coefficient is positive\footnote{For some values of $m_\Delta^2$ and $M_\Delta^2$, the diffusion coefficient develops a nonvanishing imaginary part. In this case, we consider the sign of the real part of the diffusion coefficient. This does not affect our conclusions as this happens only in regions which are anyway excluded by convexity restoration, see the following discussion.}. 
For $\xi^{-}\to0$ and $\sqrt{k^2+m_\Delta^2}-2\mu\to0$,  Eqs.~(\ref{eq:Diff}) and (\ref{eq:Adv_E_m}), respectively, develop a pole. 
These poles affect a ``self-healing'' property of the Wetterich equations~\cite{,Zorbach:2024zjx} and lead to convexity restoration -- for a proof and a detailed explanation of the relation between the poles and convexity restoration, see Ref.~\cite{litim2006convexityeffectiveactionfunctional}. 
The poles put constraints on the values of $m_\Delta^2$ and $M_\Delta^2$ that can be attained during the FRG flow.
The parts of the $(m_\Delta^2, M_\Delta^2)$ plane which are excluded for a given set of $\mu$ and $k$ are shown by the hashed regions in Figs.~\ref{fig:regionplot}. 
The blue-hashed regions are excluded by the pole in Eq.~\eqref{eq:Diff} and the orange-hashed regions by the pole in Eq.~\eqref{eq:Adv_E_m}\footnote{There are actually also regions where $\xi^+ \to 0$, but they are located inside the blue-hashed regions and not displayed in Figs.~\ref{fig:regionplot}.}. 
In the white non-hashed regions,  the diffusion coefficient is negative.
The magenta-colored points in Figs.~\ref{fig:regionplot} are the  $(m_\Delta^2, M_\Delta^2)$ values calculated from the numerically determined regularized function $u(\Delta)$ shown in Fig.~\ref{fig:comparison} (a).
Even though all values of the derivative of the potential in the UV ($k= \Lambda_{\text{UV}}=1\,\text{GeV}$) lie inside the positive-diffusion region, the plots for $k=0.39$ GeV and $k=0.075$ GeV show that some values of $u$ enter the region of negative diffusion during the FRG flow. 
These correspond to the red-colored and orange-colored points in Fig.~\ref{fig:comparison} for the respective values of $k$.
The correlation between negative diffusion and the occurrence of oscillations becomes apparent in Figs.~\ref{fig:comparison} (c) and (d).

After having identified a negative diffusion coefficient as the origin of the oscillations, we added a hyperdiffusion term of the form~(\ref{eq:Regularization}) to Eq.~\eqref{eq:KT}.
As Figs.~\ref{fig:comparison} (a, b) show, this successfully eliminates the oscillatory behavior.
While any sufficiently large $C$ in Eq.~(\ref{eq:Regularization}) would suffice to regularize the high-$q$ modes, it should still be chosen as small as possible -- since it introduces a systematic error
and additionally makes the equation significantly stiffer and thus harder to solve. 
For more details on  the numerical implementation of the hyperdiffusion term  and how the extrapolation $C\to 0$ is performed, see Appendix~\ref{sec:appendix_B}.

\subsection{Regions of Negative Diffusion in the Regularized Phase Diagram of the QDM}
\label{sec:phasediagram}

\begin{figure}[htbp]
\begin{minipage}{0.325\textwidth}
    \centering
    \includegraphics[width=\linewidth]{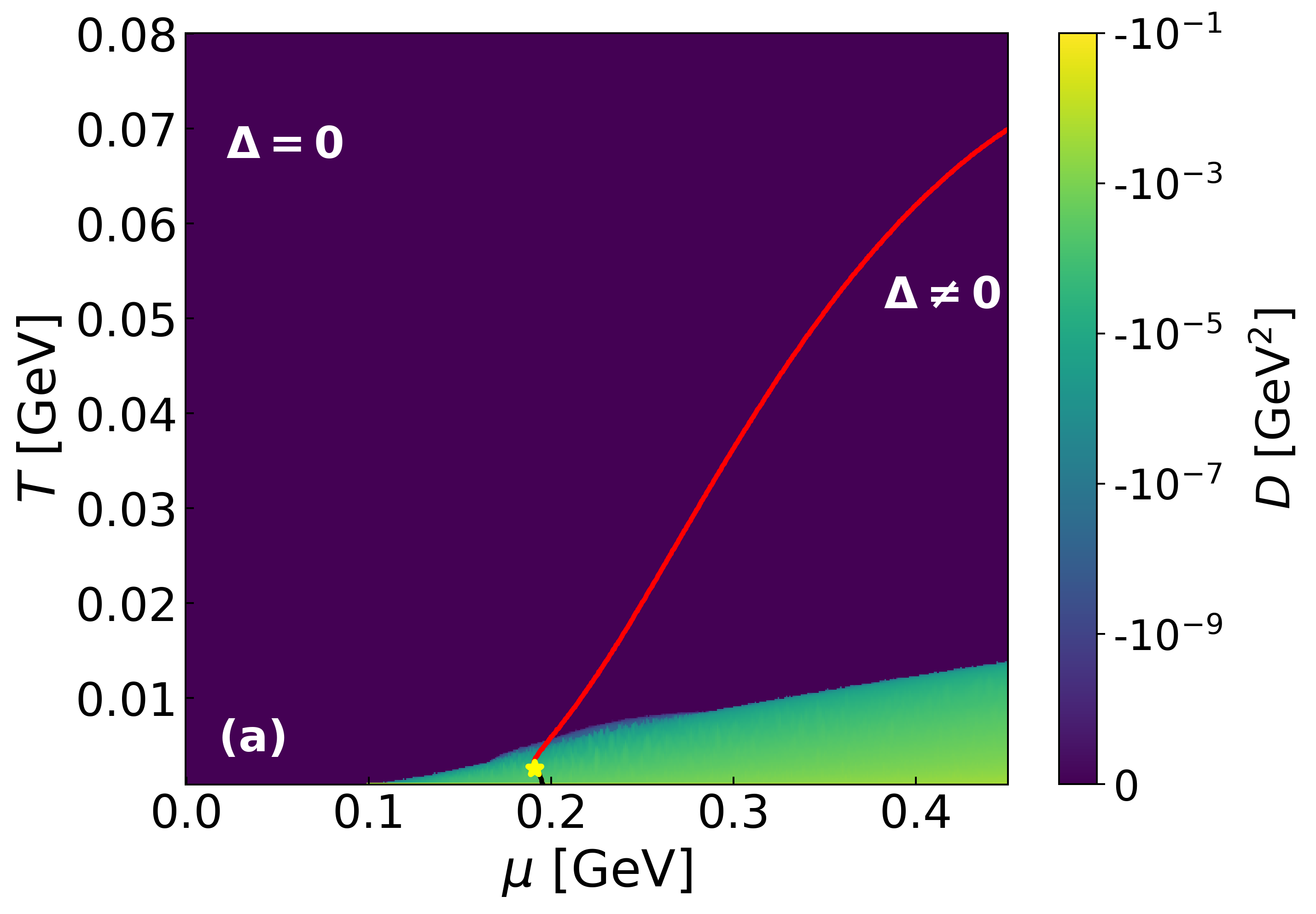}
  \end{minipage}
  \begin{minipage}{0.325\textwidth}
    \centering

    \includegraphics[width=\linewidth]{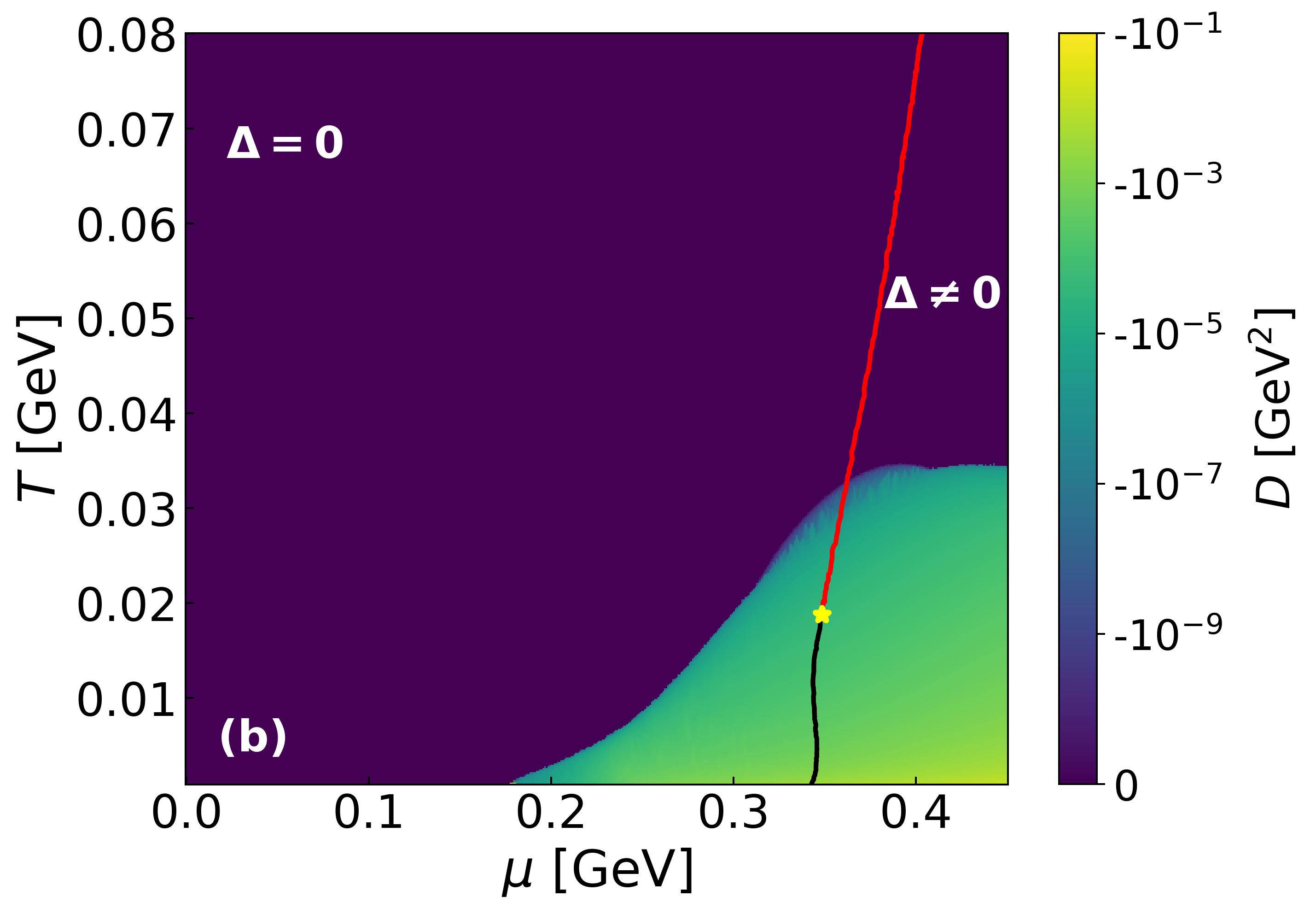}
  \end{minipage}
  \begin{minipage}{0.325\textwidth}
    \centering
    \includegraphics[width=\linewidth]{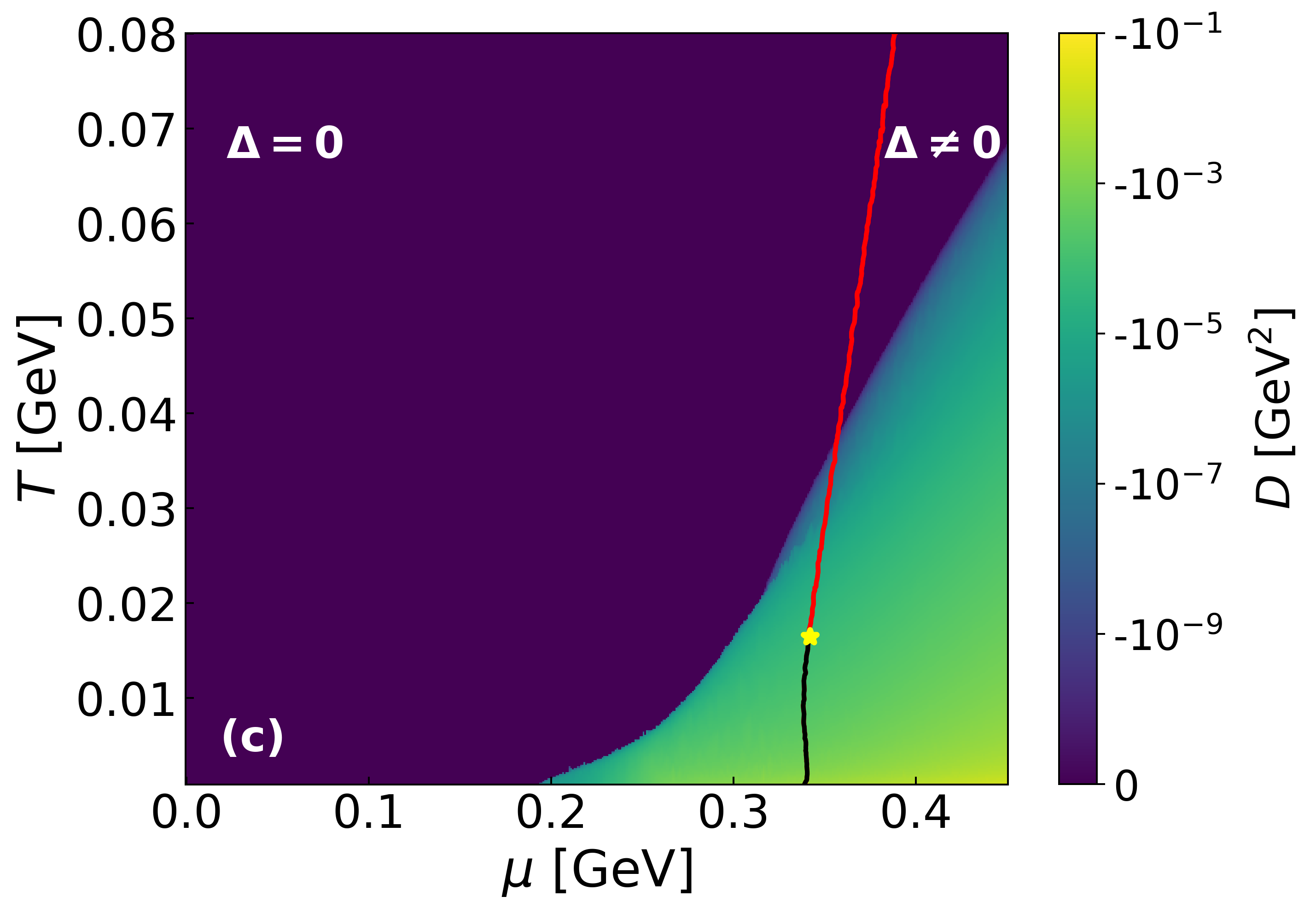}
  \end{minipage}
\caption{\label{fig:phase} Phase diagram of the QDM in the $(\mu,T)$ plane. The black (red) line shows the first-order (second-order) phase transition, separating the normal-conducting ($\Delta =0$) from the color-superconducting phase ($\Delta \neq 0$). The yellow-colored star represents the critical point separating the first- and second-order phase transition lines.
The color scale shows the most negative value of the diffusion coefficient during the flow of the derivative of the effective potential.  The parameter sets used can be found in Tab.~\ref{tab:1}, with set 1 for (a), set 2 for (b), and set 3 for (c).}
\end{figure}

We now compute the phase diagram of the QDM with the hyperdiffusion method to regularize the FRG flow in regions of negative diffusion.
The resulting phase diagram in the $(\mu, T)$ plane is shown in Fig.~\ref{fig:phase}, for the three parameter sets listed in Tab.~\ref{tab:1}.
The black (red) line shows the first-order (second-order) phase transition between the normal-conducting phase ($\Delta =0$) and the color-superconducting phase ($\Delta \neq 0$), with the yellow star representing the critical point separating the first- and second-order transition lines. 
The color scale shows the most negative value of the diffusion coefficient during the FRG flow.
The phase diagram in Fig.~\ref{fig:phase} (a) is calculated with the same parameters (set 1 in Tab.~\ref{tab:1}) as in Ref.~\cite{stoll2025nonperturbative}, where the authors chose parameters that give a diquark gap parameter of 100 MeV at a chemical potential of $\mu = 0.35\,\text{GeV}$. 
In contrast, we calculate phase diagram in Fig.~\ref{fig:phase} (b) with parameter set 2 in Tab.~\ref{tab:1}, to obtain a vacuum curvature mass of $M_\Delta^2 = 0.648 \,\mathrm{GeV}^2$, which is a more realistic value for the diquark curvature mass than the $M_\Delta^2 = 0.025\,\mathrm{GeV}^2$ obtained with parameter set 1, while getting a similar gap parameter at $\mu = 0.35\,\text{GeV}$.

To check the RG consistency~\cite{Braun:2018svj} of Fig.~\ref{fig:phase} (b), we use parameter set 3, which has an UV cutoff $\Lambda_{\text{UV}} =$ 5 GeV instead of 1 GeV, and the resulting phase diagram is shown in Fig.~\ref{fig:phase} (c).
Comparing the position of the phase-transition line in Figs.~\ref{fig:phase} (b) and (c), we conclude that RG consistency is reasonably well fulfilled. 

We observe that the stronger quark-diquark coupling in Figs.~\ref{fig:phase} (b) and (c) not only increases the size of the negative-diffusion region by extending it to higher temperatures, but also leads to more negative diffusion coefficients.
The most negative value of the diffusion coefficient across the entire phase diagram
is $D_{\mathrm{min}} =-0.0044$ GeV in Fig.~\ref{fig:phase} (a), $D_{\mathrm{min}} =-0.014$ GeV in Fig.~\ref{fig:phase} (b), and $D_{\mathrm{min}} = -0.021$ GeV in Fig.~\ref{fig:phase} (c), respectively. Correspondingly, we find stronger oscillations for parameter sets 2 and 3 than for parameter set 1 in the not-regularized FRG flow. 

These results indicate that both a stronger diquark coupling and/or enforcing RG consistency enlarge the region of negative diffusion, and hence, any physically reliable result at low temperature and high chemical potential requires a regularization scheme such as the hyperdiffusion method used here.

\section{Conclusions}
\label{sec:conclusions}

Fluid-dynamical algorithms have been shown to provide a powerful tool to solve the FRG flow equations for the effective potential~\cite{Grossi:2019urj,Koenigstein:2021syz,Sattler:2025hcg}.
The analogy to fluid dynamics allows a physically intuitive interpretation of the various terms appearing in the flow equation for the derivative of the effective potential.
For instance, one of these terms corresponds to a (nonlinear) diffusion term, which appears in analogous form in the time-honored heat equation.
In a recent work~\cite{stoll2025nonperturbative}, the FRG flow equations for the QDM were solved with such a fluid-dynamical algorithm.
The authors reported problems in the low-temperature, large-chemical potential region of the QDM phase diagram, which they attributed to numerical artifacts.
In this work, we have shown that these problems appear when the coefficient in front of the above mentioned diffusion term becomes negative. 
It is well known~\cite{Tikhonov1995} that this leads to illposedness of the PDE, and several methods to cure this have been suggested.
Here, we used the hyperdiffusion method~\cite{witelski1995shocks,miller2025shock,Maron:2008sb} to regularize the negative diffusion term.
The physical results were then extracted by extrapolating the results to the limit where the hyperdiffusion term vanishes.
In this way, we demonstrated that physically reliable results for the phase diagram of the QDM can be obtained.

In fluid dynamics, negative diffusion can sometimes be caused by truncations which insufficiently capture the nonlinear dependence of the diffusion coefficient on the fluid-dynamical fields and their gradients, and can be cured by improving the truncation to sufficiently high order. 
Therefore, it might be possible that truncations beyond the LPA (e.g., a field-dependent wave-function renormalization) could remove the negative diffusion observed here for the QDM at low temperatures and high chemical potentials. 
However, at this point we cannot exclude the possibility that negative diffusion is a genuine feature of the model.

Our results can be of particular importance for current investigations of the QCD phase diagram at nonzero quark chemical potential, where a moat regime and/or inhomogeneous phases may occur~\cite{Pisarski:2021qof,Motta:2024rvk,Fu:2024rto,pawlowski2025inhomogeneousinstabilitieshighdensityqcd}, since the FRG flow equations used to study these phenomena contain higher-order derivatives of the effective potential, which will be particularly sensitive to negative diffusion. 
Furthermore, the increased numerical stability achieved by regularization with the hyperdiffusion method allows one to flow further into the IR and to reliably extract physical observables.  

It is also worthwhile to point out that  models describing Bose-Einstein condensation or pion condensation have flow equations that share structural similarities with the QDM and QMDM and thus are also expected to develop negative diffusion in a certain parameter region~\cite{Terazaki:2024evv,Svanes_2011}. 
Therefore, the regularization method used in this paper might be helpful to regularize oscillations possibly arising in these models. 
Finally, investigating the effect of other regularization schemes remains a task for the future.

\begin{acknowledgments}
The authors thank U.~Mire, K.~Jamaly, and L.~Kiefer for fruitful discussions. 
This work is supported in part by the Deutsche Forschungsgemeinschaft (DFG, German Research Foundation) through the CRC-TR 211 ``Strong-interaction matter under extreme conditions'' -- project number
315477589-TRR 211. J.P.~acknowledges support from the Stiftung Giersch.
\end{acknowledgments}

\appendix

\section{Numerical solution of the FRG flow equation}\label{sec:appendix_A}
When solving the FRG flow equation numerically, we solve Eq.~\eqref{eq:KT} directly, i.e., we do not employ the decomposed form \eqref{eq:decomp_Q} for the diffusion term. 
Then, for the diffusion part in Eq.~\eqref{eq:KT} we apply the discretization scheme proposed in Ref.~\cite{Kurganov2000Scheme}, while for the flux part we use the Harten-Lax-van Leer-Einfeld (HLLE) algorithm (see, for example, Ref.~\cite{toro2013riemann}), where we interpolate the $u$ at the cell boundaries via a 5th-order weighted Essentially Non-Oscillatory (wENO) interpolation~\cite{Shu2009wENO}. 
For the $\partial_\Delta^4 u$ derivative, we use a 4th-order central-difference stencil. The numerical integration is performed with the \texttt{LSODA} solver from the \texttt{solve\_ivp} method in the \texttt{scipy} library~\cite{2020SciPy-NMeth}.

For our calculations, we discretize $u$ on a uniform grid with a grid spacing of $2\,\text{MeV}$, with $\Delta$ in the interval [0,$\Delta_{\mathrm{max}}$] which can be found in Tab.~\ref{tab:1}.
The boundary conditions are taken as in Ref.~\cite{Koenigstein:2021syz}. 
For the initial condition, the effective potential is taken to have the form
\begin{equation}
	U_\mathrm{UV} = \frac{m^2}{2}\Delta^2 + \frac{\lambda}{4}\Delta^4\;,
\end{equation}
where the values of the model parameters for each set can be found in Tab.~\ref{tab:1}. 
Note that we have to add a $\mu$-dependent contribution to the mass $m^2\to m^2 +4\mu^2$ in parameter set 1, in order to match the initial condition of Ref.~\cite{stoll2025nonperturbative}.

For the IR cutoff, we choose $\Lambda_{\text{IR}}=0.075\,\text{GeV}$, but for high $\mu$ and low $T$, the strong convexity restoration in this region can cause the solver to terminate at earlier RG times, similar to the numerical difficulties encountered in the QMM. 
This is of no consequence to the diffusion coefficient shown in Fig.~\ref{fig:phase}, as it reaches its highest values before reaching the IR.
The physical point was determined by finding the root of
\begin{equation}
    0 = -4\mu^2\Delta + u(\Delta)\;,
\end{equation}
the first term arising from the temporal component of the kinetic term in Eq.~(\ref{eq:Lagrangian}) when evaluated on a constant background.

All data used in this work can be found in \cite{poeplau_2026_22031211}.
\begin{table}[htbp]
	\begin{tabular}{|l|l|l|l|l|l|}
    \hline
		Test case&  $m^2$ & $\lambda$\hspace{0.5cm} & $h\hspace{0.4cm}$& $\Lambda_{\mathrm{UV}}$ & $\Delta_{\mathrm{max}}$\\ \hline
		1   & 0.0575+4$\mu^2\hspace{0.1cm}$ & 0.0   &$1.0$ & 1.0   & 2.0 \\ \hline
		2    & 0.94  & 0.1    &$3.0$    & 1.0 & 2.0\\ \hline
		3    & 6.05  & 1.0    &  $2.8$ & 5.0  & 5.0\\ \hline
	\end{tabular}
	\caption{\label{tab:1} Parameter sets used in this work. All variables are in units of GeV, or appropriate powers of it. }
\end{table}

\section{Numerical treatment of the hyperdiffusion term}
\label{sec:appendix_B}
At the current RG time step, the diffusion coefficient~(\ref{eq:Diff_fkt}) is first numerically determined over the entire $\Delta$ grid, then the smallest negative value of $D$, $\mathrm{min}_\Delta \{ D \}$, is stored.
For the next RG time step, the coefficient $C$ appearing in the hyperdiffusion term in Eq.~\eqref{eq:Regularization} is chosen as
\begin{equation}\label{eq:chosen_regularization}
	C = c\, a^2 \bar{D}\,,\quad \quad\quad
\bar{D} =
\begin{cases}
-\,\min\limits_{\Delta}\{D\}\,, & \min\limits_{\Delta}\{D\} < 0\,,\\[6pt]
\hspace{0.05\textwidth} 0\,, & \min\limits_{\Delta}\{D\} \ge 0\,.
\end{cases}
\end{equation}
with the grid spacing $a$ and $c>0$ being a number of order 1, which we take to be sufficiently large to suppress the oscillations. 
Choosing the factor $\bar{D}$ in the above manner, the hyperdiffusion term is zero if the diffusion coefficient is positive semi-definite over the entire grid.
A nonvanishing hyperdiffusion term is thus only applied when necessary, i.e., when negative diffusion occurs.
In this way, the systematic error introduced by the hyperdiffusion term stays minimal, since its magnitude is proportional to the negative diffusion coefficient which it aims to regularize.

For the regularization choice~(\ref{eq:chosen_regularization}), the limit $C\to0$ can be realized in two different ways. 
Either, one reduces the grid spacing $a\to0$ or one takes the limit $c\to0$. 
The value of $c$ cannot be set arbitrarily small, as a too small value no longer regularizes the modes with the highest $q$, causing the oscillations to reappear. 
In this work, we only consider the limit $c\to0$, as reducing the grid spacing substantially increases the numerical costs.
We realize the limit $c\to 0$ by calculating a chosen observable for different values of $c$ and performing a least-squares fit of the form 
\begin{equation}\label{eq:Fit}
	\partial_\Delta u(c) = \alpha + \beta \sqrt{c}
\end{equation}
to the data. 
As observable we choose the curvature mass $M_\Delta^2$, which we compare for the parameter sets in Tab.~\ref{tab:1} at $T=0.01$ GeV and $\mu=0.35$ GeV. 
Since the numerical values of $M_\Delta^2$ differ significantly between these sets, we instead consider their relative deviation from the limit $c\to0$ in Eq.~\eqref{eq:Fit}
\begin{equation}
    \delta(c) = \left|\frac{M_\Delta^2(c)-\alpha}{\alpha}\right|\;.
\end{equation}

\begin{figure}
	\centering
	\includegraphics[width = 0.7\paperwidth]{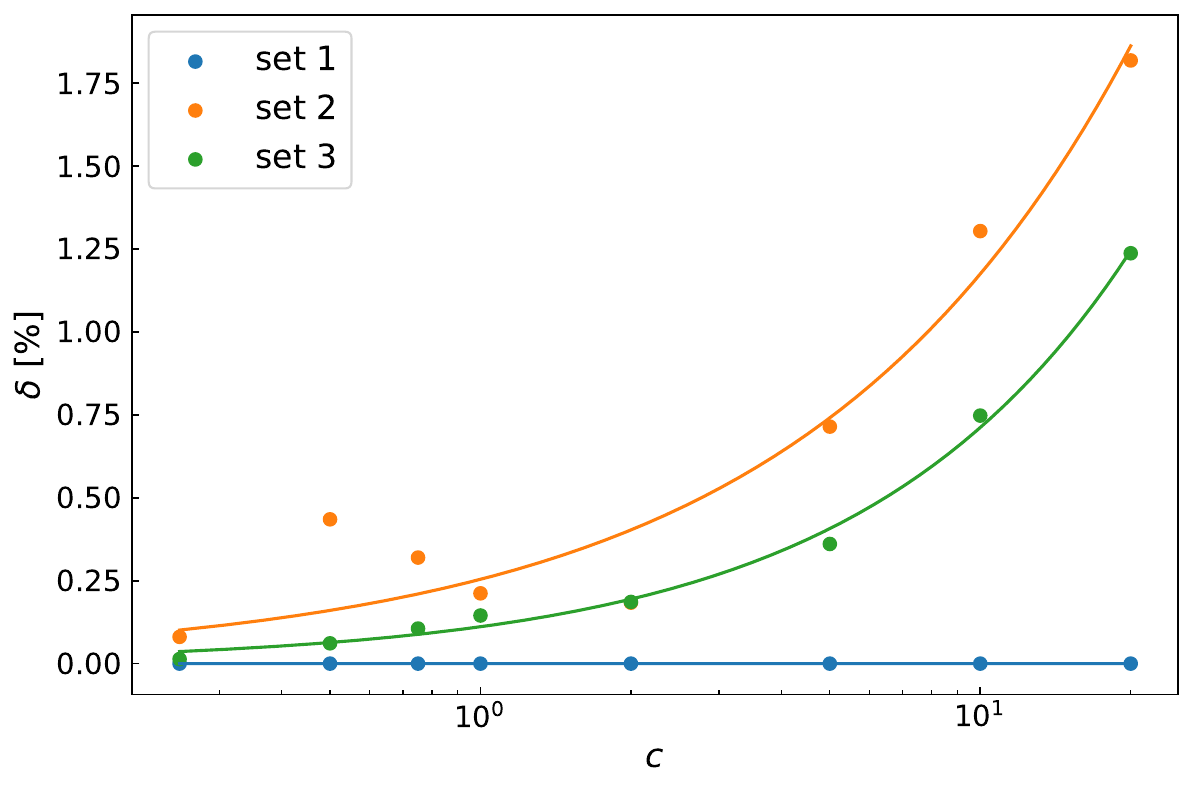}
	\caption{\label{fig:convergence} Relative convergence of $M_\Delta^2$ as the regularization term is taken to zero, for the parameter sets in Tab.~\ref{tab:1} at $T=0.01$ GeV and $\mu=0.35$ GeV. The points show the result of the FRG flow for different values of $c$, while the lines show the least-squares fit from Eq.~(\ref{eq:Fit}).}
\end{figure}

In Fig.~\ref{fig:convergence}, we show the behaviour of $\delta$ as a function of $c$. 
The symbols represent the solution of Eq.~\eqref{eq:KT} and the solid lines are the corresponding fit~\eqref{eq:Fit}. 
The $c$ dependence of the blue curve is almost negligible, since at the particular values of $T$ and $\mu$ the absolute value of the negative diffusion coefficient is comparatively small.
As can be seen, a reasonably small error in the convergence of $M_\Delta^2$ is reached already for values of $c$ of the order of one.
Therefore, if not stated differently, we set $c = 1$.

For the value of the order parameter $\Delta$, the effect of $c$ was observed to be almost negligible. 
We are therefore optimistic that the systematic error, introduced by the regularization, can be systematically removed at least for some observables, making this regularization a viable choice.


\bibliography{Ref}

\end{document}